\documentclass[12pt,a4paper]{article}
\usepackage{epsfig}
\begin{document}
\title{ Single production of the gauge boson $W$ via polarized  $e^{-}\gamma$
collisions in the littlest Higgs model }
\author{Chong-Xing Yue, Feng Zhang, Li-Na Wang, Li Zhou\\
{\small  Department of Physics, Liaoning Normal University, Dalian
116029, China}\thanks{E-mail:cxyue@lnnu.edu.cn}\\}
\date{\today}

\maketitle
\begin{abstract}
In the framework of the littlest Higgs($LH$) model, we study
single production of the standard model($SM$) gauge boson
$W_{s}^{-}$ and the heavy gauge boson $W_{H}^{-}$ via polarized
$e^{-}\gamma$ collisions. We find that the corrections of the $LH$
model to the cross section $\sigma(W_{s}^{-})$ might be observed
only for the scale parameter $f\leq 1.5TeV$ and the mixing
parameter $c'\geq 0.4$ in future high energy linear $e^{+}e^{-}$
collider($LC$) experiment with the center-of-mass($CM$) energy
$\sqrt{S}=500GeV$ and a yearly integrated luminosity
$\pounds=100fb^{-1}$. However, with a suitably chosen polarization
of the initial electron and position beams, the possible signals
of the heavy gauge boson $W_{H}^{-}$ can be easily detected via
$e^{-}\gamma$ collisions in future $LC$ experiment with
$\sqrt{S}=3TeV$ and $\pounds=500fb^{-1}$.

 \vspace{1cm}


\end{abstract}

\newpage
\noindent{\bf I. Introduction}

The high energy linear $e^{+}e^{-}$ collider $(LC)$ has a large
potential for the discovery of new particles$[1]$. Due to its
rather clean environment, the $LC$ will be perfectly suited for
precise analysis of physics beyond the standard model $(SM)$ as
well as for testing the $SM$ with an unprecedented  accuracy. A
unique feature of the $LC$ is that  it can be transformed to
$\gamma\gamma$ collisions and $e^{-}\gamma$ collisions with the
photon beams generated by the backward Compton scattering of the
initial electron and laser beams. Their effective luminosity and
energy are expected to be comparable to those of the $LC$. In some
scenarios, they are the best instrument for the discovery of
signatures of new physics.

The $e^{-}\gamma$ collisions can produce particles which are
kinematically not accessible in the  $e^{+}e^{-} $ collisions at
the same collider$[2]$. For example, for the process
$e^{-}\gamma\rightarrow AB$ with light particle A and new heavy
particle B, the discovery limits can be much higher than in other
reactions. The $e^{-}\gamma$ collisions can uniquely be identified
due to the net $(-1)$ charge in the final state, the process
$e^{-}\gamma\rightarrow AB$ offers the possibility for both new
physics discovery and precision measurements. Thus, the
$e^{-}\gamma$ collisions is particularly suitable for studying
heavy gauge boson production. In Ref.$[3]$, we have studied single
production of the heavy gauge bosons $B_{H},Z_{H}$ and $W_{H}$
predicted by the littlest  Higgs ($LH$) model$[4]$ via the
unpolarized $e^{-}\gamma$ collisions and discuss the possibility
of detecting these new particles in future $LC$ experiments. We
find that, in wide range of the parameter space preferred by the
electroweak precision data, the gauge bosons $B_{H}$ and $Z_{H}$
should be observed via detecting the $e^{-}l^{+}l^{-}$ signal.
However, the gauge boson $W_{H}$ can not be detected via the
process $e^{-}\gamma\rightarrow \nu_{e} W^{-}_{H}$.

An important tool of the $LC$ is using of the polarized beams. One
expects that a high polarization degree between $80\%$ and $90\%$
can be reached$[1,5]$. Beam polarization is not only useful for a
possible reduction of the background, but might also serve as a
possible tool to disentangle different contributions to the signal
and to directly analyze the interaction structure of new physics.
Beam polarization of the electron and position beams would lead to
a substantial enhancement of the production cross section for some
specific processes with a suitably chosen polarization
configuration. Furthermore, it has been shown that the
polarization of the initial laser beam and the electron beam will
significantly affect the photon spectrum in $e\gamma$  or
$\gamma\gamma$ collisions. Thus, a more detailed study of single
production of heavy gauge bosons in $e^{-}\gamma$ collisions with
polarized beams is needed. In this paper, we first consider the
corrections of the $LH$ model to single production of the $SM$
gauge boson$W^{-}_{s}$ via $e^{-}\gamma$ collisions and discuss
the possibility of detecting the virtual correction effects in
future $LC$ experiment with the center-of-mass($CM$) energy
$\sqrt{S}=500GeV$ and a yearly integrated luminosity
$\pounds=100fb^{-1}$. We find that the corrections might be
observed only for the scale parameter $f\leq 1.5TeV$ and the
mixing parameter $c'\geq 0.4$. Then, we study single production of
the heavy gauge boson $W_{H}^{-}$ via polarized $e^{-}\gamma$
collisions at the $LC$ with the $CM$ energy $\sqrt{S}=3TeV$ and
$\pounds=500fb^{-1}$. We find that, for a suitably chosen
polarization configuration $(p_{e},p_{\overline{e}})=(-0.8,0.6)$
and $(-0.8,-0.6)$, the production cross section of the process
$e^{-}\gamma\rightarrow \nu_{e} W^{-}_{H}$ can be significantly
enhanced and the possible signals of the heavy gauge boson
$W_{H}^{-}$ can be easily observed in wide range of the parameter
space preferred by the electroweak precision data.

Little Higgs theory$[6]$ was recently proposed as a possible
mechanism of electroweak symmetry breaking $(EWSB)$ and is a
compelling possibility for physics beyond the $SM$. The key
feature of this kind of theory is that the Higgs boson is a
pseudo-Goldstone boson of a spontaneously  broken approximate
global symmetry. So far, a number of specific models have been
proposed. The $LH$ model$[4]$ is one of the simplest and
phenomenologically viable models, which has all essential features
of the little Higgs theory. So, in the rest of this paper, we will
give our results in detail in the context of the $LH$ model.

In sec.II, we generally give the formula of the helicity
amplitudes and the cross section for the process
$e^{-}\gamma\rightarrow \nu_{e} W^{-}$, in which $W^{-}$ is the
$SM$ gauge boson $W^{-}_{s}$ or the heavy gauge boson $W_{H}^{-}$
predicted by the $LH$ model. The relative corrections of the $LH$
model to single production of the $SM$ gauge boson $W^{-}_{s}$ in
the $LC$ with $\sqrt{S}=500GeV$ and $\pounds=100fb^{-1}$  are
calculated in sec.III. The cross section of single production for
the heavy gauge boson $W^{-}_{H}$ via polarized $e^{-}\gamma$
collisions at the $LC$ with $\sqrt{S}=3TeV$ and
$\pounds=500fb^{-1}$ are calculated in sec. IV. The observability
of $W_{H}$ are also studied in this section. Section V contains
our conclusions.

\noindent{\bf II. The helicity amplitudes and the cross section
for the process $e^{-}\gamma\rightarrow \nu_{e} W^{-}$
\hspace*{0.4cm} in the $LH$ model}

At the tree-level, there are two Feynaman  diagrams contributing
to the process $e^{-}\gamma\rightarrow \nu_{e} W^{-}$ for single
production of the gauge boson $W^{-}$, as shown in Fig.1. The
s-channel diagram is induced by the gauge couplings of the gauge
bosons $\gamma$ and $W^{-}$ to fermions, while the t-channel
diagram involves a triple gauge boson coupling making this process
suitable for testing the non-Abelian gauge structure of the
theory.

\begin{figure}[htb]
\vspace{-6cm}
\begin{center}
\epsfig{file=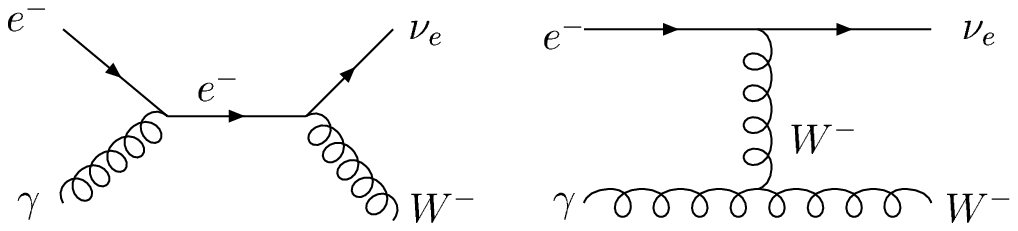,width=650pt,height=800pt} \vspace{-20cm}
\hspace{-0.5cm} \caption{Feynman diagrams for the process $
e^{-}\gamma\rightarrow\nu_{e}W^{-}$.} \label{ee}
\end{center}
\end{figure}

The $LH$ model$[4]$ consists of a non-linear $\sigma$ model with a
global $SU(5)$ symmetry and a locally gauged symmetry
$SU(2)_{1}\times U(1)_{1}\times SU(2)_{2}\times U(1)_{2}$. The
global $SU(5)$ symmetry is broken down to its subgroup $SO(5)$ by
a vacuum condensate $f\sim \Lambda_{s}/ 4\pi\sim TeV$, which
results in fourteen massless Goldstone bosons. Four of these
massless Goldstone bosons are eaten by the $SM$ gauge bosons, so
that the locally gauged symmetry $SU(2)_{1}\times U(1)_{1}\times
SU(2)_{2}\times U(1)_{2}$ is broken down to its diagonal subgroup
$ SU(2)\times U(1)$, identified as the $SM$ electroweak gauge
group. This breaking scenario gives rise to the new gauge bosons
$W_{H}^{\pm}$. In the $LH$ model, the couplings of the gauge boson
W to ordinary particles, which are related our calculation, can be
written as$[7]$:
\begin{equation}
g_{L}^{W\nu_{e} e}=\frac{ie}{\sqrt{2} S_{W}}A,\hspace{0.5cm}
g_{R}^{W\nu_{e}e}=0,\hspace{0.5cm} g^{\gamma WW}= -eB,
\end{equation}
where $S_{W}=\sin\theta_{W}$, $\theta_{W}$ is the Weinberg angle.
The coupling constant $B$ is always equal to 1 for the $SM$ gauge
boson $W_{s}$ and the heavy gauge boson $W_{H}$, while the
coupling constant A is equal to
$1-\frac{\nu^{2}}{2f^{2}}c^{2}(c^{2}-s^{2})$ and $-\frac{c}{s}$
for the gauge bosons $W_{s}$ and $W_{H}$, respectively.
$\nu=246GeV$ is the electroweak scale and $c$ $(s=\sqrt{1-c^{2}})$
is the mixing parameter between $SU(2)_{1}$ and $SU(2)_{2}$ gauge
bosons.

In our calculation, we will neglect the initial state electron
mass and think that the initial state electron beams are
longitudinally polarized beams. Thus, the negative and positive
polarized electrons coincide with their left- and right- chirality
states, respectively. In the $SM$ and the $LH$ model, the helicity
of the incoming $e^{-}$ is fixed by the massless neutrino
$\nu_{e}$, which implies that the right-handed electron has no
contributions to the cross section of the process
$e^{-}\gamma\rightarrow \nu_{e} W^{-}$. In this case, the single
production process of the charged gauge boson $W^{-}$ can be
written as:
\begin{equation}
e_{L}^{-}(p_{e})+\gamma(k)\rightarrow\nu_{eL}(p)+W^{-}(p_{W}).
\end{equation}
The helicity amplitudes of this process can be written as$[8,9]$:
\begin{equation}
M_{\lambda\lambda^{'}}=\frac{ie^{2}A}{\sqrt{2}S_{W}}
\overline{u_{\nu}}(p)T_{\mu\nu}u_{e}(p_{e})
\varepsilon^{\nu}(k,\lambda)\varepsilon
_{W}^{\mu^{*}}(p_{W},\lambda^{'}),
\end{equation}
where $\varepsilon^{\nu}(k,\lambda)$  and
$\varepsilon^{\mu}_{W}(p_{W},\lambda^{'})$ are the polarization
vectors of the initial state photon and final state $W^{-}$,
respectively. The helicities $\lambda=\pm1$ and
$\lambda^{'}=\pm1,0.$ The tensor $T_{\mu\nu}$ is the sum of the
two terms corresponding to the s- and t- channel diagrams:
\begin{equation}
T_{\mu\nu}=\frac{\gamma_{\mu}(\not p_{e}+\not
k)\gamma_{\nu}}{\hat{S}}+
\frac{1}{t-M_{W}^{2}}[\gamma^{\rho}-\frac{(\not p-\not
q)(k-p_{W})^{\rho}}
{M_{W}^{2}}]\Gamma_{\nu\mu\rho} \\
\end{equation}
with
\begin{equation}
\Gamma_{\nu\mu\rho}=2p_{W\nu}g_{\mu\rho}+2k_{\mu}g_{\nu\rho}-(p_{W}+k)_{\rho}g_{\mu\nu}
,
\end{equation}
where $t=(p-p_{e})^{2}$, $\hat{S}$ is the $CM$ energy of polarized
$e^{-}\gamma$ collisions, and $ M_{W}$ is the mass of the gauge
boson W. The first term of $Eq.(4)$ comes from the s-channel
diagram and the second term comes from the t-channel diagram.
Since there is pure vector-axial current and the charged gauge
bosons $ W^{\pm}$ has only left-handed couplings to the fermions,
the s-channel diagram has non-zero contributions to the helicity
amplitudes only for $\lambda=-1$ and $\lambda^{'}=-1,0.$

Using $Eqs.(3)-(5)$, we can write the helicity amplitudes as:
\begin{eqnarray}
M_{11}&=&\frac{C_{1}C_{2}\sqrt{\hat{S}}}{t-M_{W}^{2}}(\cos\theta-1)
\sin\frac{\theta}{2} , \\
M_{1-1}&=&\frac{C_{1}C_{2}}{t-M_{W}^{2}}[2|\vec{p}|\sin\theta-
\sqrt{\hat{S}}(1+\cos\theta)\sin\frac{\theta}{2}] , \\
M_{10}&=&\frac{C_{2}C_{3}}{t-M_{W}^{2}}[|\vec{p}|(E_{W}+|\vec{p}|)-
\sqrt{\hat{S}}E_{W}]\sin\theta\sin\frac{\theta}{2} , \\
M_{-11}&=&\frac{C_{1}C_{2}\sqrt{\hat{S}}}{t-M_{W}^{2}}[\sin\theta\cos
\frac{\theta}{2}-(1+\cos\theta)\sin\frac{\theta}{2}] , \\
M_{-1-1}&=&-C_{1}C_{2}\{\frac{2}{\sqrt{\hat{S}}}\sin\frac{\theta}{2}+
\frac{1}{t-M_{W}^{2}}[2|\vec{p}|\sin\theta+\sqrt{\hat{S}}\sin\theta\cos
\frac{\theta}{2}\nonumber \\
&&-\sqrt{\hat{S}}(\cos\theta-1)\sin\frac{\theta}{2}]\},  \\
M_{-10}&=&-C_{2}C_{3}\{\frac{(E_{W}+|\vec{p}|)\cos\frac{\theta}{2}}
{\sqrt{\hat{S}}}+\frac{1}{t-M_{W}^{2}}[\sqrt{\hat{S}}(|\vec{p}|+E_{W}
\cos\theta)\cos\frac{\theta}{2}\nonumber\\
&&-[|\vec{p}|(E_{W}+|\vec{p}|)\sin\theta-\sqrt{\hat{S}}E_{W}\sin\theta]
\sin\frac{\theta}{2}]\}
\end{eqnarray}
with
\begin{equation}
C_{1}=\frac{ie^{2}}{2S_{W}}\sqrt{\sqrt{\hat{S}}E_{\nu}}A,\hspace{0.5cm}
C_{2}=1+\frac{|\vec{p}|}{E_{\nu}},\hspace{0.5cm}
C_{3}=\frac{ie^{2}}{\sqrt{2}M_{W}S_{W}}\sqrt{\sqrt{\hat{S}}E_{\nu}}A.
\end{equation}
 In above equations, we have taken the electron momentum to be
along the z-axis and the $\theta$ represents the angle between the
electron momentum and the W momentum.

From above discussions, we can see that the chirality cross
sections $\hat{\sigma}_{RL}$ and $\hat{\sigma}_{RR}$ vanish
identically. Then the polarized cross section of the process
$e^{-}_{L}(p_{e})+\gamma(k)\rightarrow\nu_{e_{L}}(p)+W^{-}(p_{W})$
can be written as:
\begin{equation}
\hat{\sigma}(p_{e},\hat{\xi_{2}},\hat{S})=\frac{1}{4}(1-p_{e})[(1-\xi_{2})
\hat{\sigma}_{LL}+(1+\xi_{2})\hat{\sigma}_{LR}],
\end{equation}
where the Strokes parameter $\xi_{2}$ is given by [10]:
\begin{equation}
\xi_{2}=\frac{1}{D}\{p_{\bar{e}}r\xi[1+(1-x)(2r-1)^{2}]-p_{L}(2r-1)[\frac{1}{1-x}+(1-x)]\}
\end{equation}
\begin{equation}
D=\frac{1}{1-x}+1-x-4r(1-r)-p_{\bar{e}}p_{L}r\xi(2r-1)(2-x),
\end{equation}
where $p_{e}$ and $p_{\bar{e}}$ are the degrees of the
longitudinal electron and positron polarization, respectively.
$p_{L}$ is the laser photon circular polarization.
$r=\frac{x}{\xi(1-x)}$. In our calculation, we will take
$\xi=4.8$, $p_{L}=1$, and $x_{max}=0.83$ as in Ref.$[11]$.

The effective cross section $\sigma(S)$ for the subprocess
$e^{-}\gamma\rightarrow \nu_{e} W^{-}$ in a $LC$ with the $CM$
energy $\sqrt{S}$, where the positron beam with the degree of
longitudinal polarization $p_{\bar{e}}$ is converted into the
backscattered photon beam, is given by$[10]$:
\begin{equation}
\sigma(p_{e},p_{\bar{e}},S)=\int_{M_{W}^{2}/S}^{0.83}dx
F(x,p_{\bar{e}})\hat{\sigma}(p_{e},\xi_{2},\hat{S}),
\end{equation}
in which $\hat{S}=xS$, the backscattered photon distribution
function $F(x,p_{\bar{e}})$ for $p_{L}=1$ and $\xi=4.8$ is:
\begin{equation}
F(x,p_{\bar{e}})=\frac{1}{1.83+0.15p_{\bar{e}}}[\frac{1}{1-x}+1-x-
4r(1-r)-4.8p_{\bar{e}}r(2r-1)(2-x)].
\end{equation}

In the following sections, we will use these formula to calculate
the effective cross section of the sub-process
$e^{-}\gamma\rightarrow \nu_{e} W^{-}$ for $W^{-}=W_{s}^{-}$ or
$W_{H}^{-}$.

\noindent{\bf III. Single production of the $SM$ gauge boson
$W_{s}^{-}$ via polarized $e^{-}\gamma$ collisions \hspace*{0.8cm}
in the $LH$ model }

The process $e^{-}\gamma\rightarrow \nu_{e} W^{-}$ is one of the
interesting processes for $e^{-}\gamma$ collisions which can
uniquely be identified due to the net(-1) charge in the final
state. It offers the possibility for both new physics discovery
and precision measurements. Thus, studying this process in some
popular specific models beyond the $SM$ is very interesting.
Single production of the $SM$ gauge boson $W_{s}^{-}$ via the
process $e^{-}\gamma\rightarrow \nu_{e} W^{-}_{s}$ receives two
kinds of additional contributions in the $LH$ model. One comes
from the correction terms to the tree-level $W\nu_{e}e$ coupling
shown as $Eq.(1)$ and the other comes from the modification of the
relation between the $SM$ parameters and the precision electroweak
input parameters.

In the $LH$ model, the relation between the Fermi coupling
constant $G_{F}$, the $SM$ gauge boson $W_{s}$ mass $M_{W}$ and
the fine structure constant $\alpha$ can be written as$[12]$:
\begin{equation}
\frac{G_{F}}{\sqrt{2}}=\frac{\pi\alpha}{2M_{W}^{2}S_{W}^{2}}
[1-c^{2}(c^{2}-s^{2})\frac{\nu^{2}}{f^{2}}+2c^{4}\frac{\nu^{2}}
{f^{2}}-\frac{5}{4}(c^{'2}-s^{'2})\frac{\nu^{2}}{f^{2}}].
\end{equation}
So we have
\begin{equation}
\frac{e^{2}}{S_{W}^{2}}=\frac{4\sqrt{2}G_{F}M^{2}_{W}}{1-c^{2}
(c^{2}-s^{2})\frac{\nu^{2}}{f^{2}}+2c^{4}\frac{\nu^{2}}{f^{2}}-\frac{5}{4}
(c^{'2}-s^{'2})\frac{\nu^{2}}{f^{2}}}.
\end{equation}
In the following numerical estimation, we will take
$G_{F}=1.16137\times 10^{-5} GeV^{-2}$ and $M_{W}=80.45GeV[13]$ as
input parameters and use them to represent other $SM$ input
parameters.

Except for the $SM$ input parameters, there are three free
parameters: the mixing parameters $c$, $c'$, and the scale
parameter $f$ in the expression of the relative correction
parameter $R=\delta\sigma(W_{s})/\sigma^{SM} (W_{s})$ with
$\delta\sigma(W_{s})=\sigma^{LH}(W_{s})-\sigma^{SM}(W_{s})$.
$\sigma^{LH}(W_{s})$ and $\sigma^{SM}(W_{s})$ are the production
cross sections of the process $e^{-}\gamma\rightarrow \nu_{e}
W_{s}$ predicted by the $LH$ model and the $SM$, respectively. The
value of the relative correction parameter $R$ is insensitive to
the degrees of the electron and positron polarization and the $CM$
energy $\sqrt{S}$. Thus, in our numerical calculation of this
section, we do not consider the polarization of the initial states
and take $\sqrt{S}=500GeV$.
\begin{figure}[htb]
\vspace{-0.5cm}
\begin{center}
\epsfig{file=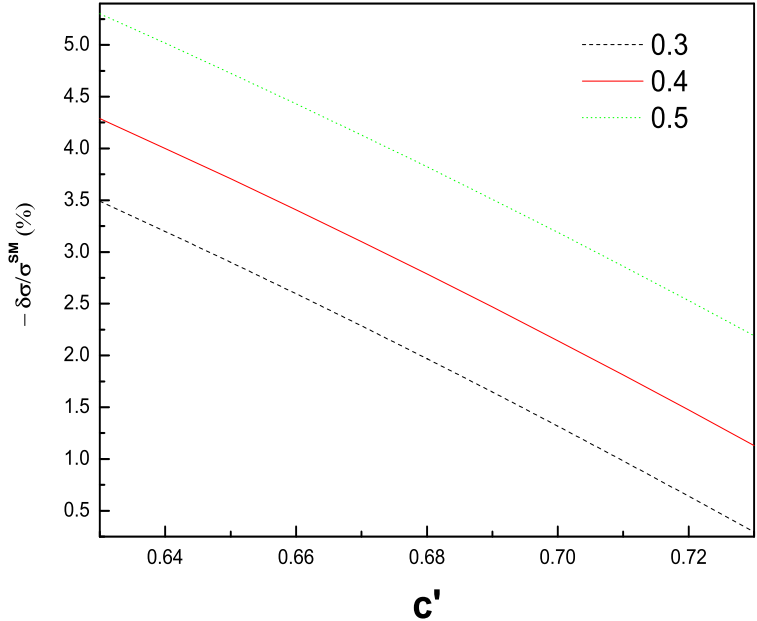,width=320pt,height=280pt} \vspace{-1cm}
\hspace{-0.5cm} \caption{The relative correction parameter $R$ as
a function of the mixing parameter $c'$ \hspace*{1.8cm} for
$f=1TeV$ and three values of the mixing parameter $c$.} \label{ee}
\end{center}
\end{figure}

In the $LH$ model, the custodial $ SU(2)$ global symmetry is
explicitly broken, which can generate large contributions to some
electroweak observables. If one assumes that the $SM$ fermions are
charged only under $U(1)_{1}$, then global fit to the electroweak
precision data produces rather severe constraints on the free
parameters of the $LH$ model[12,14]. However, if the $SM$ fermions
are charged under $U(1)_{1}\times U(1)_{2}$, the constraints
become relaxed. The scale parameter $f=1\sim 2TeV$ is allowed for
the mixing parameters $c$ and $c'$ in the ranges of $0\sim 0.5$
and $0.62\sim 0.73$, respectively[15]. Taking into account the
electroweak  precision constraints on the $LH$ model, our
numerical results are shown in Fig.2, in which we plot the
relative correction parameter $R$ as a function of the mixing
parameter $c'$ for $f=1TeV$ and three values of the mixing
parameter $c$. From Fig.2 one can see that the contributions of
the $LH$ model to the process $e^{-}\gamma\rightarrow \nu_{e}
W^{-}_{s}$ is very small. In most of the parameter space allowed
by the electroweak precision data, the absolute value of $R$ is
smaller than $5\%$.

\begin{figure}[htb]
\vspace{-1cm}
\begin{center}
\epsfig{file=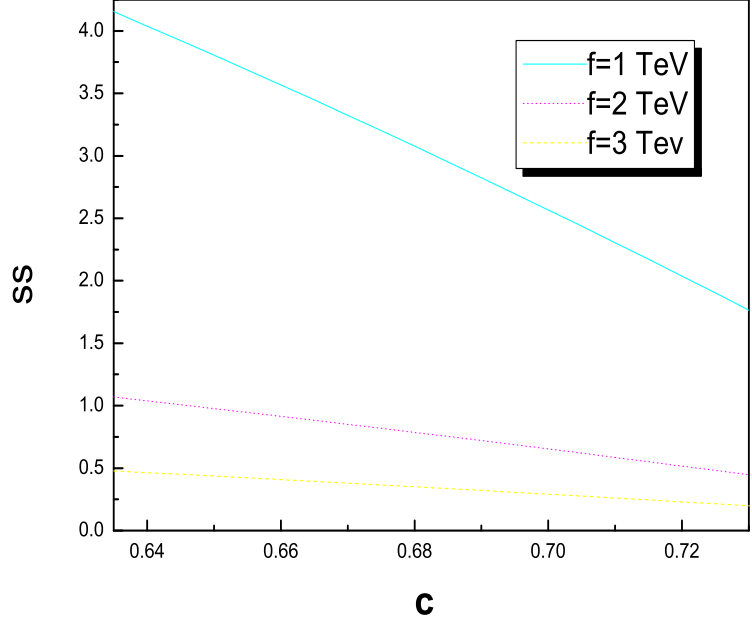,width=320pt,height=280pt} \vspace{-1cm}
\hspace{-0.5cm} \caption{The value of SS as a function of the
mixing parameter $c'$ for $c=0.3$ and \hspace*{1.8cm} and three
values of the scale parameter $f$.} \label{ee}
\end{center}
\end{figure}

In order to see whether the correction effects of the $LH$ model
on the process $e^{-}\gamma\rightarrow\nu_{e}W_{s}^{-}$ can be
observed in the near future $LC$ experiment with $\sqrt{S}=500GeV$
and $\pounds=100fb^{-1}$, we define the statistical significance
(SS) of the signal as:
\begin{equation}
SS=\frac{|\sigma^{LH}(W_{s})-\sigma^{SM}(W_{s})|}{\sqrt{\sigma^{SM}}}\sqrt{\pounds
}.
\end{equation}

In order to ensure that the events are well within the detector
range, we demand that the angles of all detectable final particles
with the beam pipe are smaller than 15$^{0}$. With this
requirement, we assumed that 60$\%$ of the produced $W_{s}$ can be
properly reconstructed. In Fig.3, we plot SS as a function of the
mixing parameter $c'$ for $c=0.4$ and three values of the scale
parameter $f$. From Fig.3 one can see that, for $f=1TeV$ and
$0.62\leq c'\leq0.71$, the value of SS is larger than 2, while,
for $f\geq 1.5TeV$, its value is smaller than 2 in most of the
parameter space. Thus, the correction effects of the $LH$ model on
the process $e^{-}\gamma\rightarrow\nu_{e}W_{s}^{-}$ might be
detected at the $LC$ with $\sqrt{S}=500GeV$ and
$\pounds=100fb^{-1}$ only for $f\leq1.5TeV$ and $c'\geq0.4$.

 \noindent{\bf IV. Single production of the heavy
gauge boson $W_{H}^{-}$ via polarized $e^{-}\gamma$ collisions}

From above discussions, we can see that the single production
cross section $\sigma(W_{H})$ of the heavy gauge boson $W^{-}_{H}$
via the process $e^{-}\gamma\rightarrow \nu_{e} W^{-}_{H}$
dependents on the mixing parameter $c'$ only through the relation
between the $SM$ parameters and the precision electroweak input
parameters as shown in $Eq.(18)$. Thus, the $c'$ dependence of
$\sigma (W_{H})$ is very weak and we can take the fixed value for
the mixing parameter $c'$. Taking into account the electroweak
precision constraints on the $LH$ model, we assume $c'=0.65$, $
c=0.1\sim 0.5$, and $M_{W_{H}}=1\sim 3 TeV$ in our following
calculation.

Figure 4 shows the dependence of the production cross section
$\sigma(W_{H})$ on the mixing parameter $c$ for $\sqrt{S}=3TeV$,
$c'=0.65$, and four values of the $W_{H}$ mass $M_{W_{H}}$. To see
the effects of the electron and positron beam polarization on
$\sigma (W_{H})$, we have chosen the different polarization
configuration $(p_{e},p_{\bar{e}})=(0.8,0.6), (0.8,-0.6),
(-0.8,0.6), (-0.8,-0.6 )$, and $(0,0)$ in Fig.4. One can see from
Fig.4 that the production cross section $\sigma(W_{H})$ is indeed
sensitive to the polarization of electron and positron beams. For
a suitably chosen polarization configuration, the value of
$\sigma(W_{H})$ can be significantly enhanced. For example, for
$(p_{e},p_{\bar{e}})=(-0.8,0.6)$, and $(-0.8,-0.6)$, the values of
$\sigma(W_{H})$ are larger than those for
$(p_{e},p_{\bar{e}})=(0,0)$ in all of the parameter space. If we
take $0.1\leq c\leq 0.5$ and $1TeV\leq M_{W_{H}}\leq 2.5TeV$,
which is allowed by the electroweak precision constraints, then
the single production section $\sigma(W_{H})$ of the heavy gauge
boson $W_{H}$ via polarized $e^{-}\gamma$ collisions in the future
$LC$ with $\sqrt{S}=3TeV$ are in the ranges of $60fb\sim 1.4\times
10^{-2}fb$, $72.9fb\sim 3.8\times 10^{-3}fb$, and $36.3fb\sim
3.9\times 10^{-3}fb$ for $(p_{e},p_{\bar{e}})=(-0.8,0.6),
(-0.8,-0.6)$, and $(0,0)$, respectively. If we assume that the
yearly integrated luminosity of the future $LC$ experiment with
$\sqrt{S}=3TeV$ is $\pounds=500fb^{-1}$, then there are several
tens and up to thousand $W_{H}\nu_{e}$ events to be generated per
year.

\begin{figure}[htb]
\vspace{0cm}
  \centering
   \includegraphics[width=3.3in]{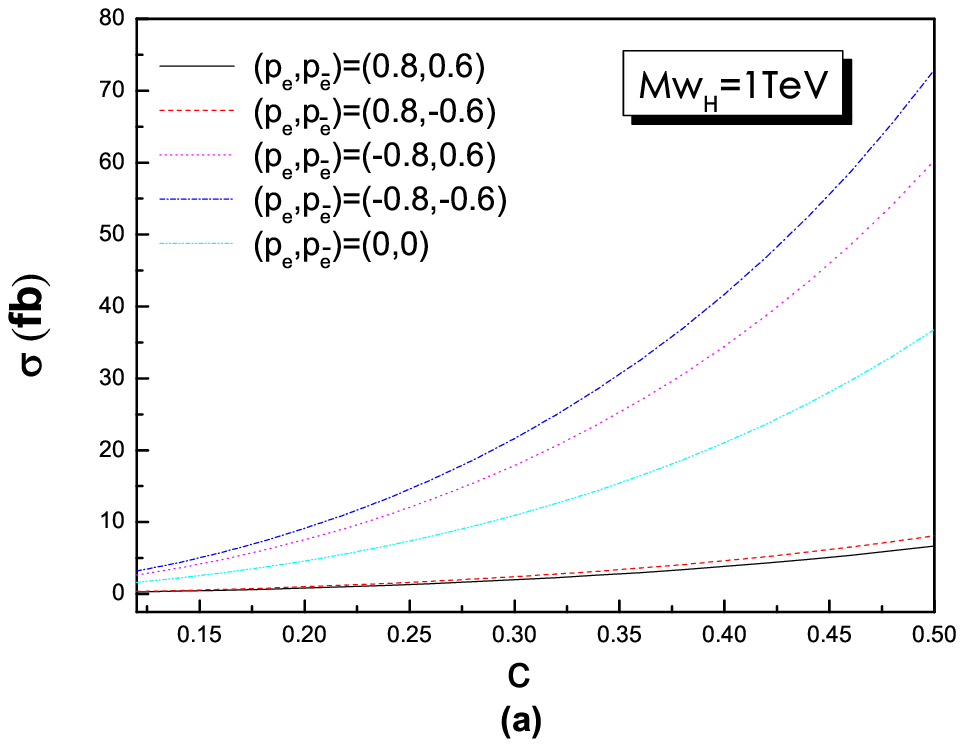}
    \hspace{-0.4in}
   \includegraphics[width=3.3in]{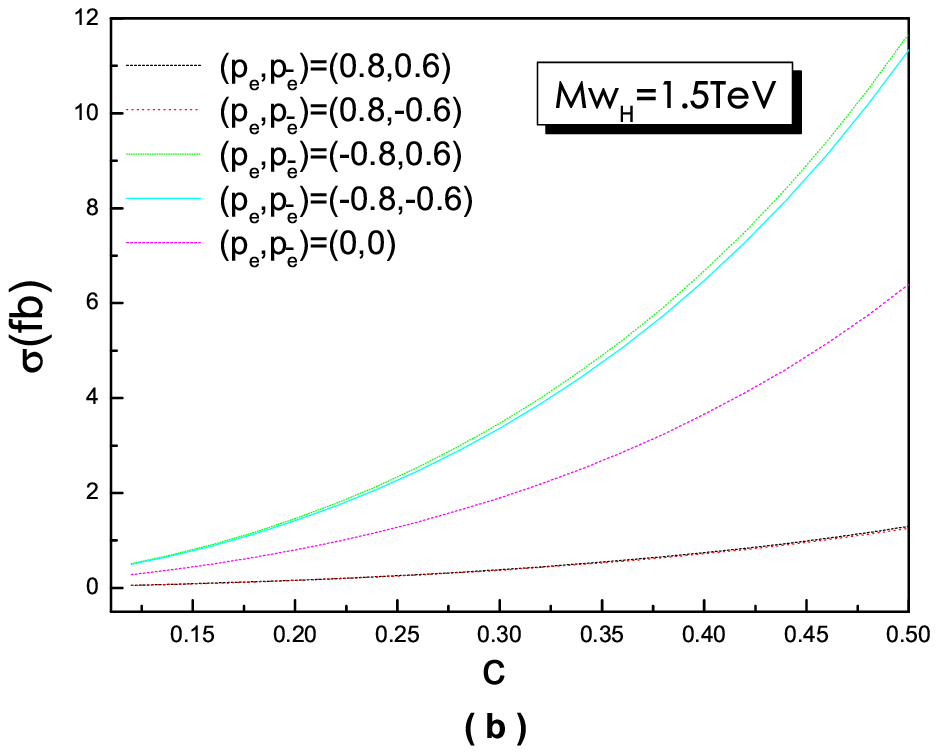}
\end{figure}

\begin{figure}[htb]
\vspace{-1cm}
  \centering
   \includegraphics[width=3.3in]{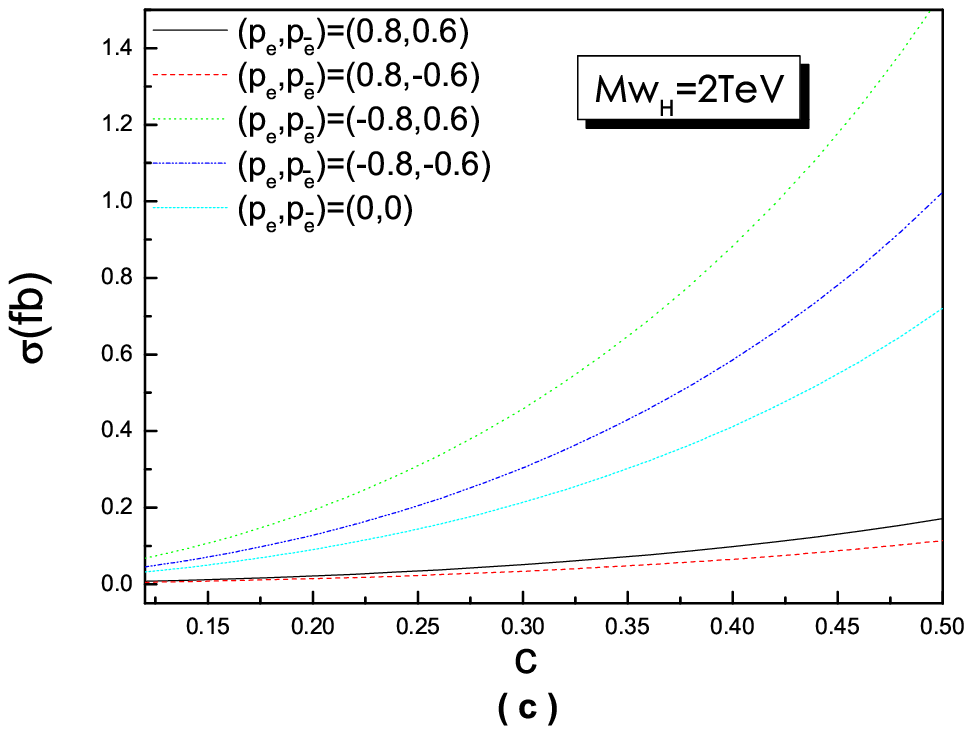}
    \hspace{-0.4in}
   \includegraphics[width=3.3in]{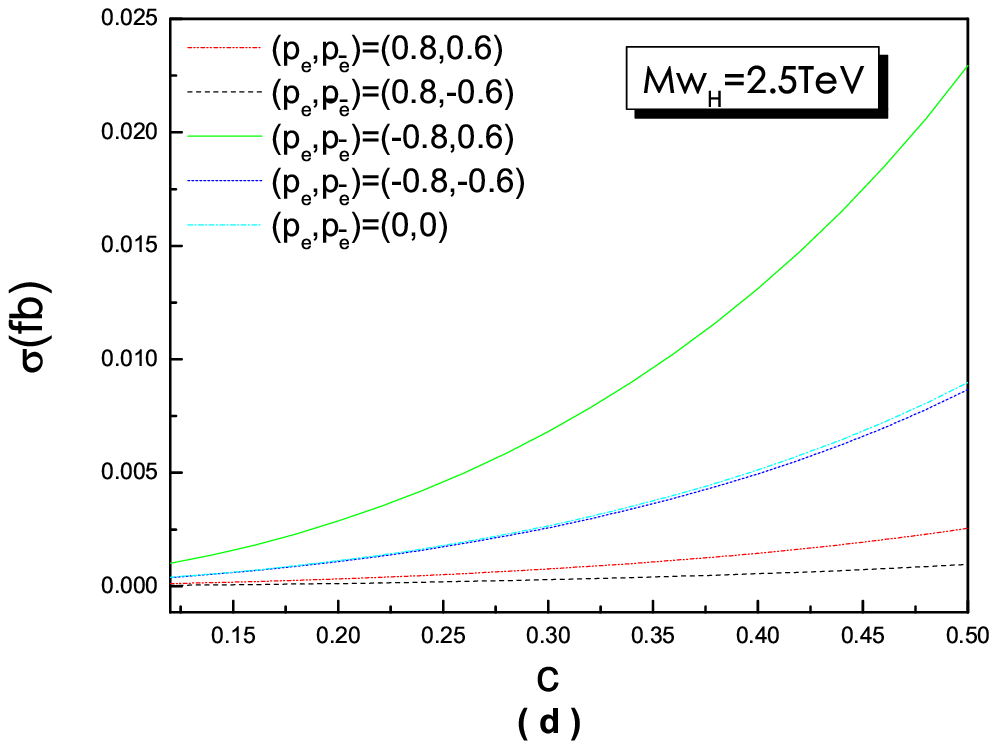}
  \caption{The cross section $\sigma(W_{H})$ as a function of the mixing parameter
  $c$ for \hspace*{1.8cm} different
  values of the $W_{H}$ mass $M_{W_{H}}$ and different polarization
   configurations.}
\end{figure}

In Fig.4 we have taken that the value of the degree $P_{L}$ of the
laser-beam polarization equals to 1. Certainly, we can also take
$P_{L}=-1$. In this case, the value of the production cross
section $\sigma(W_{H})$ is different from that for the same
polarization configuration with $P_{L}=1$. However, the conclusion
that a suitably polarization configuration of the initial electron
and positron beams can enhance the single production cross section
$\sigma(W_{H})$ is not changed.

In general, the heavy gauge bosons are likely to be discovered via
their decay products. The decay channels $W_{H}^{\pm}\rightarrow
l^{\pm}\nu$ can manifest itself via events that contain an
isolated charged lepton and missing energy. In this case, the
signal of single production of the heavy gauge boson $W_{H}^{-}$
via $e^{-}\gamma$ collisions should be an isolated charged lepton
associated with large missing energy. For the hadron decay
channels $W^{\pm}_{H}\rightarrow qq'$, the signal is a two jet
event associated with large missing energy. In the narrow width
approximation, the number of the $l^{-}\nu\nu_{e}[2j+\nu_{e}]$
events can be approximately written as
$N_{W_{H}}=\pounds\sigma(W_{H})\times B_{r}(W^{-}_{H}\rightarrow
l^{-}\nu) [B_{r}(W^{-}_{H}\rightarrow qq')] $. The branching ratio
$B_{r}(W^{-}_{H}\rightarrow l^{-}\nu[qq'])$ can be easily
estimated using the formula given by Ref.$[16]$. The most serious
backgrounds for the $l^{-}\nu\nu_{e}$ signal come from the $SM$
processes $e^{-}\gamma\rightarrow e^{-}Z\rightarrow e^{-}\nu\nu$
and $e^{-}\gamma\rightarrow W_{s}^{-}\nu_{e}\rightarrow
l^{-}\nu\nu_{e}$. The scattered electron in the process
$e^{-}\gamma\rightarrow e^{-}Z$ has almost same energy
$E_{e}\approx\frac{\sqrt{S}}{2}$ for $\sqrt{S}\gg M_{Z}$. Thus,
the process $e^{-}\gamma\rightarrow e^{-}Z$ could be easily
distinguished from the signal$[8,9]$. Furthermore, the cross
section for this process  decreases as $\sqrt{S}$ increasing,
while the cross section for the process $e^{-}\gamma\rightarrow
\nu_{e} W^{-}_{s}$ is approaching a constant at high energies. So,
the most serious background process is $e^{-}\gamma\rightarrow
\nu_{e} W^{-}_{s}\rightarrow l^{-}\nu\nu_{e}$.
\begin{figure}[htb]
\vspace{-0.5cm}
  \centering
   \includegraphics[width=3.3in]{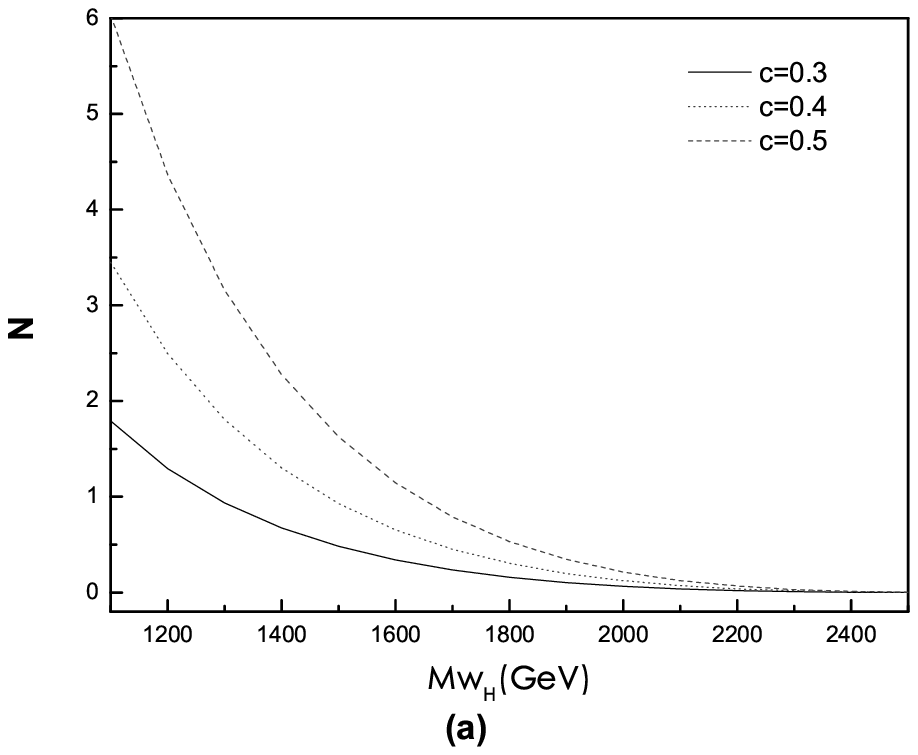}
    \hspace{-0.4in}
   \includegraphics[width=3.3in]{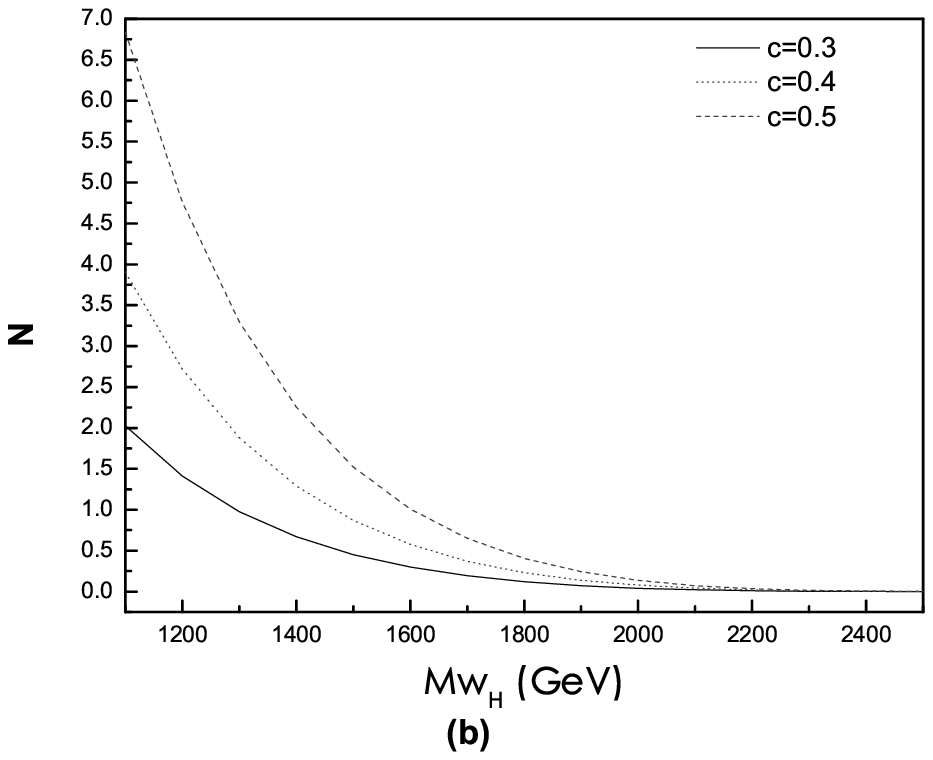}
\caption{The ratio $N$ as a function of $M_{W_{H}}$ for
 three values of the mixing parameter $c$. \hspace*{1.8cm} The
polarization of  the initial state beams are taken as
$(p_{e},p_{\bar{e}})=(-0.8,0.6)$ \hspace*{1.9cm} and $(-0.8,-0.6)$
in Fig.5(a) and Fig.5(b), respectively.}
\end{figure}

To compare the signal with background and discuss the possibility
of detecting the heavy gauge boson $W_{H}^{-}$, we calculate the
ratio of signal over square root of the background $(N=N_{W_{H}}/
\sqrt{B})$ in the parameter space of the $LH$ model preferred by
the electroweak precision data, in which
$B=\pounds\sigma(W_{s}^{-})\times B_{r}(W_{s}^{-}\rightarrow
l^{-}\nu)$ for the lepton channel $W_{H}^{-}\rightarrow l^{-}\nu$
. The dependence of $N$ on the $W_{H}$ mass $M_{W_{H}}$ is shown
in Fig.5 for $c'=0.65$ and three values of the mixing parameter
$c$. We have taken the polarization of the electron and positron
beams as $(p_{e},p_{\bar{e}})=(-0.8,0.6)$, and $(-0.8,-0.6)$ in
Fig.5(a) and Fig.5(b), respectively. One can see from Fig.5 that
the values of $N$ increase as the mixing parameter $c$ increasing
and the $W_{H}$ mass $M_{W_{H}}$ decreasing. With reasonable
values of the free parameters, the value of $N$ is larger than 2.
This means that, at least, the $W_{H}$ signal can be separated
from its $SM$ background at $2\sigma$ confidence level. Thus, the
$W_{H}$ signal might be observed via detecting the
$l^{-}\nu\nu_{e}$ event.

\begin{figure}[htb]
\vspace{-0.5cm}
\begin{center}
\epsfig{file=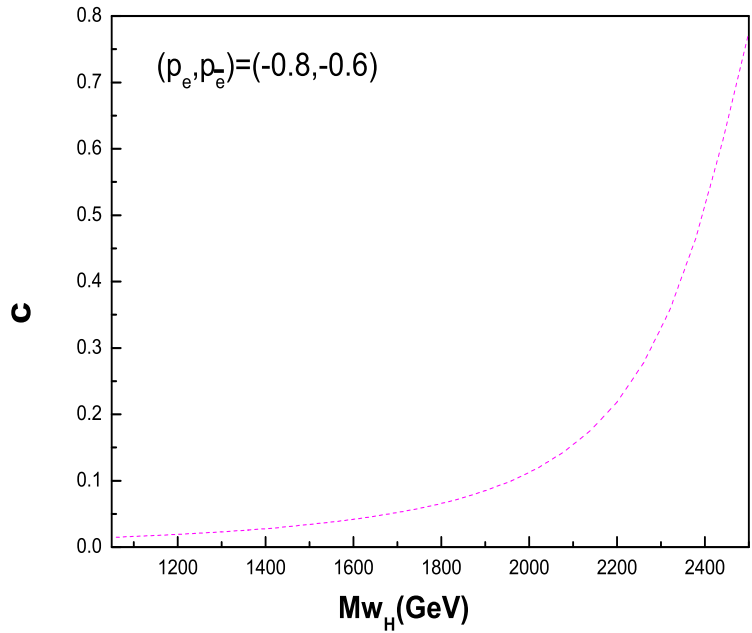,width=320pt,height=280pt} \vspace{-1cm}
\hspace{-0.5cm} \caption{ In the case of detected the gauge boson
$W^{-}_{H}$ via the $l\nu\nu_{e}$ final state, the \hspace*{1.8cm}
dependence of the mixing parameter $c$ on the $W_{H}$ mass
$M_{W_{H}}$ \hspace*{1.8cm} .} \label{ee}
\end{center}
\end{figure}

Compared to the $SM$ fermions (quarks and leptons), the $W_{H}$
mass $M_{W_{H}}$ is very large. For the $W_{H}^{-}$ decay channels
$W_{H}^{-}\rightarrow ff'$, we can neglect the fermion masses and
there is $Br(W_{H}^{-}\rightarrow l^{-}\nu)\approx
Br(W_{H}^{-}\rightarrow qq')$. For the $SM$ gauge boson $W_{s}$,
there is $Br(W_{s}^{-}\rightarrow qq')>Br(W_{s}^{-}\rightarrow
l^{-}\nu)$. Thus, the values of the ratio N for the lepton
channels $W_{H}^{-}\rightarrow l^{-}\nu$ are larger than those for
the hadron decay channels $W^{-}_{H}\rightarrow qq'$. The possible
signals of the new gauge boson $W_{H}^{-}$ should be more easy
observed via detecting the $l^{-}\nu\nu_{e}$ event than via
detecting the $qq'\nu_{e}$ event.

It is well known that a appropriate cut on the $SM$ background can
generally enhance the ratio of signal over square root of the
background. It has been shown that, with the suitably cut on the
final lepton transverse momentum and rapidity, the $SM$ background
$l^{-}\nu\nu_{e}$ can be reduced by more than one order of
magnitude$[8]$. Thus, we expect that, as long as the mixing
parameter $c>0.3$ and the $W_{H}$ mass $M_{W_{H}}< 2TeV$, the
heavy gauge bosons $W_{H}^{\pm}$ should be detected via polarized
$e\gamma$ collisions in future $LC$ experiment with
$\sqrt{S}=3TeV$ and $\pounds=500fb^{-1}$.

If we assume that the heavy gauge boson $W_{H}^{-}$ has been
observed in high-energy experiments, such as $LHC$, then we can
study the constraints on the free parameters of the $LH$ model via
considering the contributions of $W_{H}^{-}$ to the processes
$e^{-}\gamma\rightarrow l^{-}\nu\nu_{e}$ and $qq'\nu_{e}$, in
which $q$ and $q'$ are the $SM$ quarks $u$ and $d$ or $c$ and $s$.
At 95$\%$ confidence level, the constraints can be derived from
\begin{equation}
X^{2}=[\frac{\sigma(SM+W_{H})-\sigma(SM)}{\delta\sigma}]^{2}=3.84,
\end{equation}
where $\delta\sigma$ is the expected experimental uncertainty
about the corresponding cross section. In our numerical
estimation, we will assume $\delta\sigma=2\%$, the yearly
integrated luminosity $\pounds=500fb^{-1}$, and the polarization
of the electron and positron beams as ($P_{e},
P_{\overline{e}}$)=(-0.8, -0.6), and take the $l^{-}\nu\nu_{e}$
final state as an example. The constraints on the free parameters
$c$ and $M_{W_{H}}$ for $c'=0.65$ are showed in Fig.6. From this
figure, we can see that the constraints are very weak. For
example, as long as the heavy gauge boson $W_{H}$ mass $M_{W_{H}}$
is smaller than $2TeV$, the $W_{H}^{-}$ signals can be detected
via the $l^{-}\nu\nu_{e}$ final state for $c\geq0.1$. Certainly,
we can also obtain the constraints on these free parameters from
the $qq'\nu_{e}$ final state. However, the constraints are
stronger than those from the $l^{-}\nu\nu_{e}$ final state.

\noindent{\bf V. Conclusions }

To solve the so-called hierarchy or fine-tuning problem of the
$SM$, the little Higgs theory was proposed  as a kind of models of
$EWSB$ accomplished by a naturally light Higgs sector. For all of
the little Higgs models, at least two interactions are needed to
explicitly break all of the global symmetries to make the Higgs
boson as a pseudo-Goldstone boson. In general, these models
predict the existence of the new heavy gauge bosons, colored
fermions, and triplet scalars to cancel the quadratically
divergent contributions to the Higgs mass induced by the $SM$
gauge bosons, Higgs boson, and the top-quark. These new particles
might produce characteristic signatures at present or future high
energy collider experiments$[7,17]$. Studying the possible
signatures of these new particles can help to test little Higgs
theory and further to probe $EWSB$ mechanism.

The $LH$ model is one of the simplest and phenomenologically
viable models, which realizes the little Higgs idea. The high
energy $e^{-}\gamma$ collision is particularly suitable for
studying single production of the heavy gauge bosons. Thus, in the
context of the $LH$ model, we study single production of the $SM$
gauge boson $W_{s}^{-}$ and the heavy gauge boson $W_{H}^{-}$ via
polarized $e^{-}\gamma$ collisions. We find that the correction of
the $LH$ model to the production cross section of the process
$e^{-}\gamma\rightarrow \nu_{e} W^{-}_{s}$ is very small in most
of the parameter space, which is very difficult to be detected in
future $LC$ experiment with $\sqrt{S}=500GeV$ and
$\pounds=100fb^{-1}$.

Beam polarization of the electron and positron beams would lead to
a substantial enhancement of the production cross sections for
some specific processes with a suitably chosen polarization
configuration. Our numerical results show that the cross sections
of the process $e^{-}\gamma\rightarrow \nu_{e} W^{-}_{H}$ for
$(p_{e},p_{\bar{e}})=(-0.8,0.6)$ and $(-0.8,-0.6)$ are larger than
those for $(p_{e},p_{\bar{e}})=(0,0)$ in all of the parameter
pace. For $0.3<c\leq 0.5$ and $1TeV\leq M_{W_{H}}\leq 2TeV$
allowed by the electroweak precision data, the values of the
single production cross sections for the heavy gauge boson
$W_{H}^{-}$ are in the ranges of $0.46fb\sim 60fb$ and $0.3fb\sim
73fb$ for $(p_{e},p_{\bar{e}})=(-0.8,0.6)$ and $(-0.8,-0.6)$,
respectively. With reasonable values of the free parameters and a
appropriate cut on the $SM$ background $e^{-}\gamma\rightarrow
W_{s}^{-}\nu_{e}\rightarrow l^{-}\nu\nu_{e}$, the value of
$N_{W_{H}}/\sqrt{B}$ can be significantly large. Thus, the
possible signals of the heavy gauge bosons $W_{H}^{\pm}$ might be
detected via polarized $e\gamma$ collisions in future $LC$
experiment with $\sqrt{S}=3TeV$ and $\pounds=500fb^{-1}$.

\vspace{0.5cm} \noindent{\bf Acknowledgments}

This work was supported in part by Program for New Century
Excellent Talents in University(NCET-04-0290), the National
Natural Science Foundation of China under the grant No.10475037,
and the Natural Science Foundation of the Liaoning Scientific
Committee(20032101).

\end{document}